\documentclass[acmtap]{acmlarge}

\usepackage{amsmath}
\usepackage[tight,small,center]{subfigure}
\usepackage[font=small]{caption}
\usepackage{tabularx}
\usepackage{multirow}
\usepackage[pdfborder={0 0 0},pdfstartview=FitR]{hyperref}
\makeatletter
\def\url@leostyle{%
  \@ifundefined{selectfont}{\def\UrlFont{\sf}}{\def\UrlFont{\footnotesize\ttfamily}}}
\makeatother
\makeatletter
\def\hlinewd#1{%
\noalign{\ifnum0=`}\fi\hrule \@height #1 %
\futurelet\reserved@a\@xhline}
\makeatother

\newcommand{\ie}{{\it i.e. }}
\newcommand{\etal}{{\it et al. }}

\acmArticle{}
\acmYear{2012}
\acmMonth{5}

\usepackage[ruled]{algorithm2e}
\SetAlFnt{\algofont}
\SetAlCapFnt{\algofont}
\SetAlCapNameFnt{\algofont}
\SetAlCapHSkip{0pt}
\IncMargin{-\parindent}
\renewcommand{\algorithmcfname}{ALGORITHM}


\DeclareMathOperator*{\argmin}{arg\,min}

\title{On voting intentions inference from Twitter content: a case study on UK 2010 General Election}
\author{{\large\textbf{Vasileios Lampos}}\\
Intelligent Systems Laboratory\\
Computer Science Department\\
University of Bristol\\
\href{mailto:bill.lampos@bristol.ac.uk}{bill.lampos@bristol.ac.uk}
}
\begin{document}

\maketitle

\section*{Abstract}
This is a report, where preliminary work regarding the topic of voting intention inference from Social Media -- such as Twitter -- is presented. Our case study is the UK 2010 General Election and we are focusing on predicting the percentages of voting intention polls (conducted by YouGov) for the three major political parties -- Conservatives, Labours and Liberal Democrats -- during a 5-month period before the election date (May 6, 2010). We form three methodologies for extracting positive or negative sentiment from tweets, which build on each other, and then propose two supervised models for turning sentiment into voting intention percentages. Interestingly, when the content of tweets is enriched by attaching synonymous words, a significant improvement on inference performance is achieved reaching a mean absolute error of $4.34\% \pm 2.13\%$; in that case, the predictions are also shown to be statistically significant. The presented methods should be considered as work-in-progress; limitations and suggestions for future work appear in the final section of this script.

\section{Introduction}
Information generated by users and published on the Social Web has enabled a new wave of experimentation and research in the past few years. There exist several works showing that textual `observations' coming from Social Media -- such as Twitter -- could, for example, be used to nowcast events emerging in the real world \cite{Lampos2011Nowcasting}, understand general emotional figures in a population \cite{Golder2011,Lansdall-Welfare2012} and improve branding strategies \cite{Jansen2009a}.

In this report, we present a preliminary method for extracting voting intention figures from Twitter. The case study used to verify our findings is the 2010 General Election in the UK.\footnote{United Kingdom General Election 2010, \url{http://en.wikipedia.org/wiki/United_Kingdom_general_election,_2010}.} In the recent past, a few papers have been published on this topic \cite{Lui2010,tumasjan2010predicting,Gayo-Avello2011,Metaxas2011a,OConnor2010}, offering preliminary solutions or discussing the limitations that several approaches might have; we refer to and discuss them at the final section of this report.

We consider only the three major parties in the UK; namely the Conservative (\textbf{CON}), the Labour (\textbf{LAB}) and the Liberal Democrat Party (\textbf{LIBDEM} or \textbf{LIB}). Overall, we are using three techniques for extracting positive and negative sentiment from tweets and then two different methods to map this sentiment to voting intention percentages.

Ground truth is acquired by YouGov's Published Results\footnote{YouGov Archive, \url{http://labs.yougov.co.uk/publicopinion/archive/}.} and consists of 68 voting intention polls dated from January to May 2010. Polls usually refer to a pair of days (in our data set only 3 of the them are 1-day polls) and indicate an expectation for the voting intention percentage per political party. As we move closer to the election day (6th of May), they become more dense; there is a new poll published every day. Tweets are drawn from the same period of time, \ie January to May; their total number is greater than 50 million, but not all of them are used as it will become apparent in the following sections. After applying some filtering to keep tweets regarding politics, we end up with 300,000 tweets, \ie approximately 100,000 tweets per political party.

\section{Extracting positive and negative sentiment from tweets}
\label{section:positive_negative_sentiment}
The common characteristic in all three approaches is that at first we retrieve a set of tweets regarding each political party by using a handmade set of keywords (see Appendix \ref{Ap:voting_intention}). The keywords are different per party, not many (approx. 60 per party) and as simple as possible; we use the names of the top politicians in each party, the name of the party and so on. The search is case sensitive when the keyword is an 1-gram and case insensitive when it is an $n$-gram, with $n > 1$. In the latter case, we are not looking for an exact match either; we are just searching for tweets that contain all 1-grams of the target $n$-gram. Character `\#' in front of an 1-gram denotes that we are searching for a Twitter topic. Since those keywords are based mainly on names or entities, one could argue that they could also be created in an automatic manner or extracted from a repository and hence, the human involvement could become insignificant.

The three approaches build on each other; each one is a more elaborate version of its precedent. In the first approach, we are using a stemmed version of SentiWordNet 3.0 to extract positive and negative sentiment from tweets without taking into consideration the different parts of speech (verb, noun, adjective and so on). SentiWordNet is the result of automatically annotating all WordNet synsets according to their degrees of positivity, negativity and neutrality \cite{Baccianella2010,Esuli2006a}. Stemming is performed by applying Porter's algorithm \cite{porter1980} and Part-Of-Speech (\textbf{POS}) tagging is `skipped' by computing the average positive and negative sentiment weight for each stem over all possible POS that it might appear in. Stems with equal positive and negative sentiment are not considered. The positive and negative sentiment scores of a tweet are retrieved by computing the sum of positive and negative sentiment weights of the words it contains. It is noted that there might exist tweets with no words listed in SentiWordNet -- those tweets have zero positive and negative sentiment and therefore are ignored. The acronym \textbf{SnPOS} (Sentiment no POS tagging) is used to denote this approach. The motivation behind the removal of POS tagging is the assumption that Twitter language might not follow the norms of formal scripts and therefore POS taggers -- trained on more `mainstream' types of text -- might create inaccurate results. However, as we will see in practice in the following sections, this is not always the case, probably because tweets that refer to politics have a much better structure than the more casual ones.

The second approach is based on the same basic principles as SnPOS, only this time POS tagging is applied. When one stem is mapped to a particular POS more than once (this can happen due to stemming), the average positive and negative sentiment weights are assigned to it. POS tagging of tweets is carried out by using Stanford POS Tagger, which is a Java implementation of the log-linear POS taggers described in \cite{Toutanova2000,Toutanova2003}. By summing over the sentiment weights of a tweet's terms, we retrieve its sentiment score. This method is denoted as \textbf{SPOS} (Sentiment with POS tagging).

Finally, we extend SPOS, by incorporating the core word senses of WordNet \cite{Miller1995}, a semi-automatically compiled list of 5,000 terms.\footnote{Available at \url{http://wordnet.princeton.edu/wordnet/download/standoff/}.} WordNet gives a set of synonyms for each term in this list and we use those synonyms to extend the content of a tweet. That is, if a tweet contains a word listed in WordNet's core terms, the tweet is extended by attaching the synonyms of this word to it. Again, the sentiment score of a tweet is computed by summing over the sentiment weights of its POS tagged terms. This method is identified by the acronym \textbf{SPOSW} (Sentiment with POS tagging and WordNet's core terms). The main motivation behind extending the content of a tweet with synonyms is the fact that the short length of a tweet might reduce expressiveness and therefore, adding more words could enhance its semantic orientation. Furthermore, SentiWordNet does not include all English words and by attaching synonyms, we achieve to compute pseudo-sentiment for a greater number of tweets.

\section{Inferring Voting Intentions}
By applying a method from the ones described in the previous section, we retrieve the positive and negative sentiment scores for a set of $t$ tweets, which in turn has been originally extracted by using the set of keywords for a political party. The next task is to turn those scores into a percentage which will represent the voting intention for this particular party. In this section, we describe two methods that address this task.

Optionally, one could first remove tweets with almost equal positive and negative sentiments. The semantic orientation of those tweets is unclear and therefore might not always help. Later on, when we present our experimental findings, for every experimental setting we test, we are also replicating the same experiment by introducing a threshold $\delta$, which removes the top 20,000 (this is equal to approx. 6.67\% of the entire number of tweets) most `semantically unclear' tweets. Learning the optimal value of $\delta$ from data is possible, but also dependent to the existence of a larger sample.

For the remaining $m \leq t$ tweets we compute their mean positive and negative sentiment scores, say $\mu_{\text{pos}_m}$ and $\mu_{\text{neg}_m}$ respectively. The sentiment score assigned to this set of tweets is derived by simply subtracting those two quantities
\begin{equation}
\label{eq:senti_mts}
\text{Senti}(m) = \mu_{\text{pos}_{m}} - \mu_{\text{neg}_{m}}.
\end{equation}

Suppose that we have computed the sentiment scores for a set of $n$ time instances (a time instance is equal to one or two days based on the target voting intention poll) for all three parties, $\text{Senti}_{\text{CON}}^{(n)}$, $\text{Senti}_{\text{LAB}}^{(n)}$ and $\text{Senti}_{\text{LIB}}^{(n)}$. To calibrate the relation between sentiment scores and voting intention percentages, we regress each one of these vectors with the corresponding ground truth using OLS and compute the three weights $w_{\text{CON}}$, $w_{\text{LAB}}$ and $w_{\text{LIB}}$. For example,
\begin{equation}
w_{\text{CON}} = \argmin_{w}\sum_i^n \left(\text{Poll}_{\text{CON}}^{(i)} - \text{Senti}_{\text{CON}}^{(i)}w\right)^2,
\end{equation}
where $\text{Poll}_{\text{CON}}^{(i)}$ denotes the official voting intention poll percentage for time instance $i$. We do not introduce bias terms in OLS regression, as during our experimental process it became evident that they receive large values, reducing significantly the freedom of our model and causing overfitting.

In this case study, we are only considering the percentages of the three major parties. After the inference process, the results are normalised in order to represent valid percentages (having a sum equal to 1); in the special (and also rare) case, where inferences are lower than 0, negative results are thresholded to 0 and normalisation does not take place, unless the nonzero percentages sum up to a value greater than 1. Official voting intentions, however, take into consideration the existence of other political `forces' as well as people's desire not to vote and so on. To create an equivalent and comparable representation, we have normalised the official voting intention percentages as well -- based on the fact that the sum of voting intention percentages for the three major parties is on average equal to approx. 90\%, this normalisation does not change the picture much (and we can always normalise the results back to a 90\% level, if required).

Voting intention inferences are made on unseen data. Again, we have a set of $k$ time instances, which do not overlap with the $n$ time instances used for learning the calibration weights, and we first compute the sentiment scores for each party per time instance. Then, those sentiment scores are multiplied by the corresponding calibration weight; for example the inferred score for the Conservative Party will be equal to:
\begin{equation}
\text{CON}^{(k)} = \text{Senti}_{\text{CON}}^{(k)} \times w_{\text{CON}}.
\end{equation}
After computing the triplet of inferences, $\text{CON}^{(k)}$, $\text{LAB}^{(k)}$ and $\text{LIB}^{(k)}$, we normalise them so that their sum is equal to 1, for example:
\begin{equation}
\text{CON}^{(k)}_{\text{norm}} = \frac{\text{CON}^{(k)}}{\text{CON}^{(k)} + \text{LAB}^{(k)} + \text{LIB}^{(k)}},
\end{equation}
represents the $k$ normalised inferences of the voting intention for the Conservative Party. We denote the method described above as Mean Thresholded Sentiment (\textbf{MTS}).

The second method is identical to MTS, but uses a different function to compute Senti$(m)$ (Equation \ref{eq:senti_mts}). After retrieving the $m$ tweets that satisfy a threshold $\delta$, we count how many of them have a positive sentiment score that is higher than the negative and vice versa. The overall sentiment score of those tweets is then computed by
\begin{equation}
\text{Senti}(m) = \frac{\#\{\text{pos} > \text{neg}\} - \#\{\text{neg} > \text{pos}\}}{m},
\end{equation}
where $\#\{\text{pos} > \text{neg}\}$ is the number of tweets with a positive sentiment greater than the negative one and $\#\{\text{neg} > \text{pos}\}$ the number of tweets with a negative sentiment greater than the positive. We refer to this method as Dominant Sentiment Class (\textbf{DSC}). %Similarly to MTS, if Senti$(m)$ contains negative values, $c$ is set equal to $c = |\min(\text{Senti})| + \sigma(\text{Senti}(m))$, otherwise $c$ is 0.

\section{Experimental process and results}
We measure the performance of our methods using two loss functions. Primarily, we use the Mean Absolute Error (\textbf{MAE}) (and its standard deviation) between the inferred and the target voting intention percentages. This metric allows for an easier interpretation because it can be read as a percentage, \ie MAE has the same units as the inferred values. In addition, aiming to assess how good is the ranking of the political parties based on the inferred voting intention percentages, we are measuring a Mean Ranking Error (\textbf{MRE}). For each triplet of voting intention percentages, the ranking error is defined as the sum over all three parties of the distance between the correct and the inferred ranking. For example, if there exists one incorrect ranking of size 1 (where size measures the difference between a correct and an inferred position), then -- since we are dealing with 3 variables only -- either there exists one more incorrect ranking of size 1 and therefore the triplet's total error is equal to 2, or two more ranking errors of size 1 and 2, which will make the triplet's total error equal to 4. Obviously, a triplet with correct rankings has an error equal to 0 and the maximum error ($\max(\text{RE})$) per triplet is equal to 4. MRE ranges in $[0,1]$ and is computed by:
\begin{equation}
\text{MRE}(n) = \frac{1}{n*\max(\text{RE})} \sum_{i = 1}^{n} \text{RE}(i).
\end{equation}

In total, by combining the three methods for extracting positive and negative sentiment from tweets (SnPOS, SPOS and SPOSW) and the two methods for converting sentiment scores to voting intentions (MTS and DSC), we come up with six experimental setups. We run those experiments for $\delta = 0$, \ie for all tweets with a sentiment score, but also for the value of $\delta$ which removes the top 20,000 most `semantically unclear' tweets.

\begin{table}
\footnotesize
\renewcommand{\arraystretch}{1.2}
\setlength\tabcolsep{1mm}
\centering
\tbl{MAE $\pm$ MAE's standard deviation and MRE for Mean Thresholded Sentiment (\textbf{MTS}) by performing leave-one-out cross validation. $\delta$ denotes the minimum distance between the positive and negative sentiment score for a tweet in order to be considered.}{
\(\begin{tabular}{cc|ccc|c|c}
& $\delta$ & \textbf{CON} & \textbf{LAB}             & \textbf{LIBDEM}           & \textbf{All Parties}     & \textbf{MRE}\\\hline
\textbf{SnPOS} & 0      & 12.24 $\pm$ 10.66 & 10.66 $\pm$ 10.39 & 10.1  $\pm$ 11.54 & 11    $\pm$ 8.31 & 0.4559\\
\textbf{SnPOS} & 0.025  & 12.17 $\pm$ 10.63 & 10.54 $\pm$ 10.47 & 10.11 $\pm$ 11.44 & 10.94 $\pm$ 8.27 & 0.4779\\\hline
\textbf{SPOS}  & 0      & 36.06 $\pm$ 7.09  & 17.86 $\pm$ 11.29 & 15.12 $\pm$ 11.65 & 23.02 $\pm$ 5.23 & 0.9412\\
\textbf{SPOS}  & 0.0072 & 35.99 $\pm$ 7.17  & 17.88 $\pm$ 11.26 & 15.06 $\pm$ 11.65 & 22.98 $\pm$ 5.3  & 0.9412\\\hline
\textbf{SPOSW} & 0      & 4.22  $\pm$ 2.78  & 3.75  $\pm$ 2.86  & 5.1   $\pm$ 3.76  & 4.36  $\pm$ 2.13 & 0.2059\\
\textbf{SPOSW} & 0.0238 & 4.17  $\pm$ 2.88  & 3.76  $\pm$ 2.82  & 5.07  $\pm$ 3.7   & 4.34  $\pm$ 2.13 & 0.1912\\
\end{tabular}\)
}
\label{table_vi_l1out_MTS}
\end{table}

\begin{table}
\footnotesize
\renewcommand{\arraystretch}{1.2}
\setlength\tabcolsep{1mm}
\centering
\tbl{MAE $\pm$ MAE's standard deviation and MRE for Dominant Class Sentiment (\textbf{DCS}) by performing leave-one-out cross validation.}{
\(\begin{tabular}{cc|ccc|c|c}
& $\delta$ & \textbf{CON} & \textbf{LAB}             & \textbf{LIBDEM}           & \textbf{All Parties}     & \textbf{MRE}\\\hline
\textbf{SnPOS} & 0      & 9.68  $\pm$ 8.82  & 9.26 $\pm$ 8.76 & 9.34 $\pm$ 9.23 & 9.42  $\pm$ 6.66 & 0.4338\\
\textbf{SnPOS} & 0.025  & 8.23  $\pm$ 6.83  & 8.66 $\pm$ 8.34 & 9.28 $\pm$ 9.12 & 8.72  $\pm$ 6.39 & 0.375\\\hline
\textbf{SPOS}  & 0      & 11.61 $\pm$ 11.99 & 7.29 $\pm$ 5.27 & 7.9  $\pm$ 6.06 & 8.93  $\pm$ 6.21 & 0.5074\\
\textbf{SPOS}  & 0.0072 & 11.62 $\pm$ 11.89 & 7.52 $\pm$ 5.34 & 8.09 $\pm$ 6.33 & 9.08  $\pm$ 6.46 & 0.5074\\\hline
\textbf{SPOSW} & 0      & 5.5   $\pm$ 4.12  & 4.19 $\pm$ 3.35 & 5.52 $\pm$ 4.82 & 5.07  $\pm$ 3.04 & 0.2647\\
\textbf{SPOSW} & 0.0238 & 5.47  $\pm$ 3.9   & 3.95 $\pm$ 3.34 & 5.49 $\pm$ 4.49 & 4.97  $\pm$ 2.88 & 0.2279\\
\end{tabular}\)
}
\label{table_vi_l1out_DCS}
\end{table}

\begin{table}
\footnotesize
\renewcommand{\arraystretch}{1.2}
\setlength\tabcolsep{1mm}
\centering
\tbl{MAE $\pm$ MAE's standard deviation and MRE for Mean Thresholded Sentiment (\textbf{MTS}). $\delta$ denotes the minimum distance between the positive and negative sentiment score for a tweet in order to be considered.}{
\(\begin{tabular}{cc|ccc|c|c|c}
& $\delta$ & \textbf{CON} & \textbf{LAB}             & \textbf{LIBDEM}           & \textbf{All Parties}     & \textbf{MRE} & \textbf{p-value} \\\hline
\textbf{SnPOS} & 0      & 12.97 $\pm$ 10.89 & 10.56 $\pm$ 10.05 & 6.66  $\pm$ 7.08  & 10.06 $\pm$ 9.72  & 0.4038 & 0.222\\
\textbf{SnPOS} & 0.025  & 12.98 $\pm$ 10.92 & 10.51 $\pm$ 10.03 & 6.7   $\pm$ 7.19  & 10.06 $\pm$ 9.74  & 0.4038 & 0.259\\\hline
\textbf{SPOS}  & 0      & 35.03 $\pm$ 6.96  & 21.31 $\pm$ 9.4   & 12.71 $\pm$ 7.16  & 23.02 $\pm$ 12.11 & 0.9231 & 0.531\\
\textbf{SPOS}  & 0.0072 & 35.14 $\pm$ 6.82  & 21.33 $\pm$ 9.41  & 12.77 $\pm$ 7.13  & 23.08 $\pm$ 12.1  & 0.9231 & 0.527\\\hline
\textbf{SPOSW} & 0      & 4.44  $\pm$ 3.18  & 2.66  $\pm$ 1.85  & 3.74  $\pm$ 2.86  & 3.61  $\pm$ 2.76  & 0.1731 & 0.019\\
\textbf{SPOSW} & 0.0238 & 4.44  $\pm$ 3.31  & 2.65  $\pm$ 1.8   & 3.84  $\pm$ 2.71  & 3.64  $\pm$ 2.75  & 0.1346 & 0.006\\
\end{tabular}\)
}
\label{table_vi_MTS}
\end{table}

\begin{table}
\footnotesize
\renewcommand{\arraystretch}{1.2}
\setlength\tabcolsep{1mm}
\centering
\tbl{MAE $\pm$ MAE's standard deviation and MRE for Dominant Class Sentiment (\textbf{DCS}).}{
\(\begin{tabular}{cc|ccc|c|c|c}
& $\delta$ & \textbf{CON}    & \textbf{LAB}             & \textbf{LIBDEM}           & \textbf{All Parties}       & \textbf{MRE} & \textbf{p-value}\\\hline
\textbf{SnPOS}  & 0      & 10.56 $\pm$ 6.75 & 9.82 $\pm$ 9.18 & 7.69 $\pm$ 9.88 & 9.36  $\pm$ 8.68 & 0.4038 & 0.467\\
\textbf{SnPOS}  & 0.025  & 9.66  $\pm$ 6.89 & 9.46 $\pm$ 9.05 & 7.25 $\pm$ 9.04 & 8.79  $\pm$ 8.35 & 0.3654 & 0.529\\\hline
\textbf{SPOS}   & 0      & 10.63 $\pm$ 8.94 & 8.09 $\pm$ 6.37 & 6.12 $\pm$ 5.12 & 8.28  $\pm$ 7.14 & 0.3846 & 0.238\\
\textbf{SPOS}   & 0.0072 & 10.51 $\pm$ 9.14 & 7.95 $\pm$ 6.18 & 6.08 $\pm$ 5.5  & 8.18  $\pm$ 7.26 & 0.4038 & 0.149\\\hline
\textbf{SPOSW}  & 0      & 4.51  $\pm$ 3.45 & 2.87 $\pm$ 2.06 & 3.53 $\pm$ 3.29 & 3.64  $\pm$ 3.04 & 0.1154 & 0\\
\textbf{SPOSW}  & 0.0238 & 4.49  $\pm$ 3.49 & 2.46 $\pm$ 1.81 & 3.51 $\pm$ 3.14 & 3.49  $\pm$ 2.98 & 0.0962 & 0\\
\end{tabular}\)
}
\label{table_vi_DCS}
\end{table}

\begin{figure*}
    \begin{center}
    \subfigure[First 34 time instances]{\includegraphics[width=5.5in]{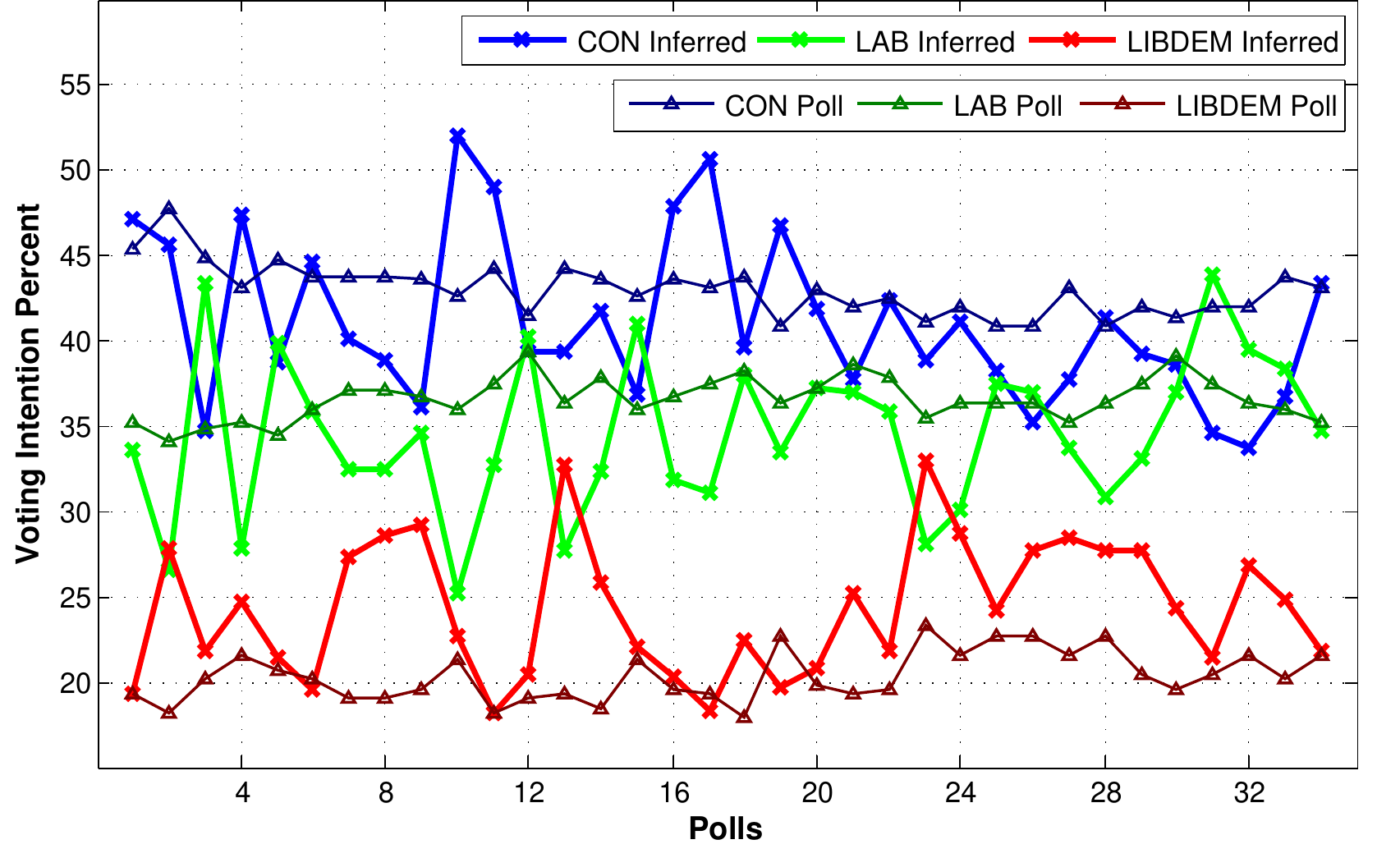}
    \label{fig_vi_l1out_MTS_SPOWS_part1}}
    \hfil
    \subfigure[Last 34 time instances]{\includegraphics[width=5.5in]{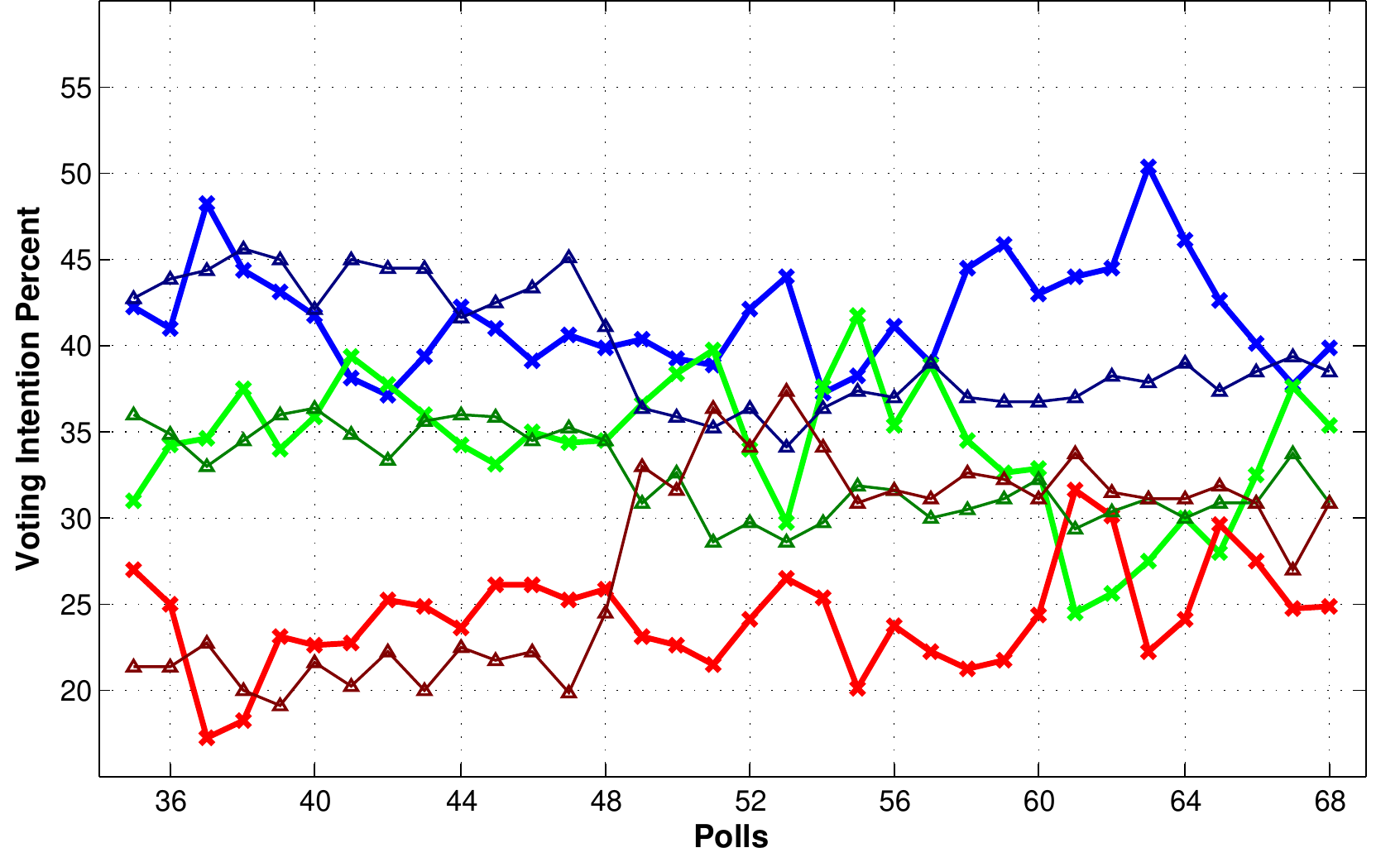}
    \label{fig_vi_l1out_MTS_SPOSW_part2}}
    \end{center}
    \caption{Leave-one-out cross validation by applying SPOSW and MTS thresholded (in two parts for a better visualisation).}
    \label{fig_vi_l1out_MTS_SPOSW}
\end{figure*}

We retrieve the first performance figures by performing leave-one-out cross validation on the aforementioned experimental setups. The performance results for MTS and DSC are presented in Tables \ref{table_vi_l1out_MTS} and \ref{table_vi_l1out_DCS} respectively. Under both methods, we see that thresholding tends -- in most occasions -- to improve the inference performance in terms of MAE and MRE. SnPOS performs significantly better than SPOS, which also fails to rank voting intentions properly under MTS. Using DCS results in a better performance on average, but does not deliver the overall best performance, which is derived by SPOSW under MTS using a thresholded data set. Figure \ref{fig_vi_l1out_MTS_SPOSW} depicts those best performing inferences against the corresponding ground truth; in this case, MAE across all parties is equal to $4.34$ with a standard deviation of $2.13$ and MRE is $0.1912$.

Since our main data set is of small size and based on the fact that the oscillations in the actual voting intention signals are not major (something that might affect the training and testing processes), we also perform a more focused testing. From the 68 polls, we use 1--30 and 47--58 for training and we perform testing on the remaining 26 ones (31--46 and 59--68). The number of tweets that we retrieve using our search terms is increasing as we get nearer to the election day. This is why the training set has been sliced into two parts, one `way' before the election day (05/01 to 24/03) and another much closer (13/04 to 25/04). In this experiment, we additionally retrieve a statistical significance indication for our inferences. To do that, we randomly permute the outcomes of MTS or DSC and come up with a randomised training and test set; we repeat this process for 1,000 times and count how many times a model that is based on randomly permuted data delivers a better inference performance than the actual one -- this fraction gives the p-value. We consider that a p-value lower than 0.05 indicates statistical significance.

\begin{figure}[!t]
\centering
\includegraphics[width=6in]{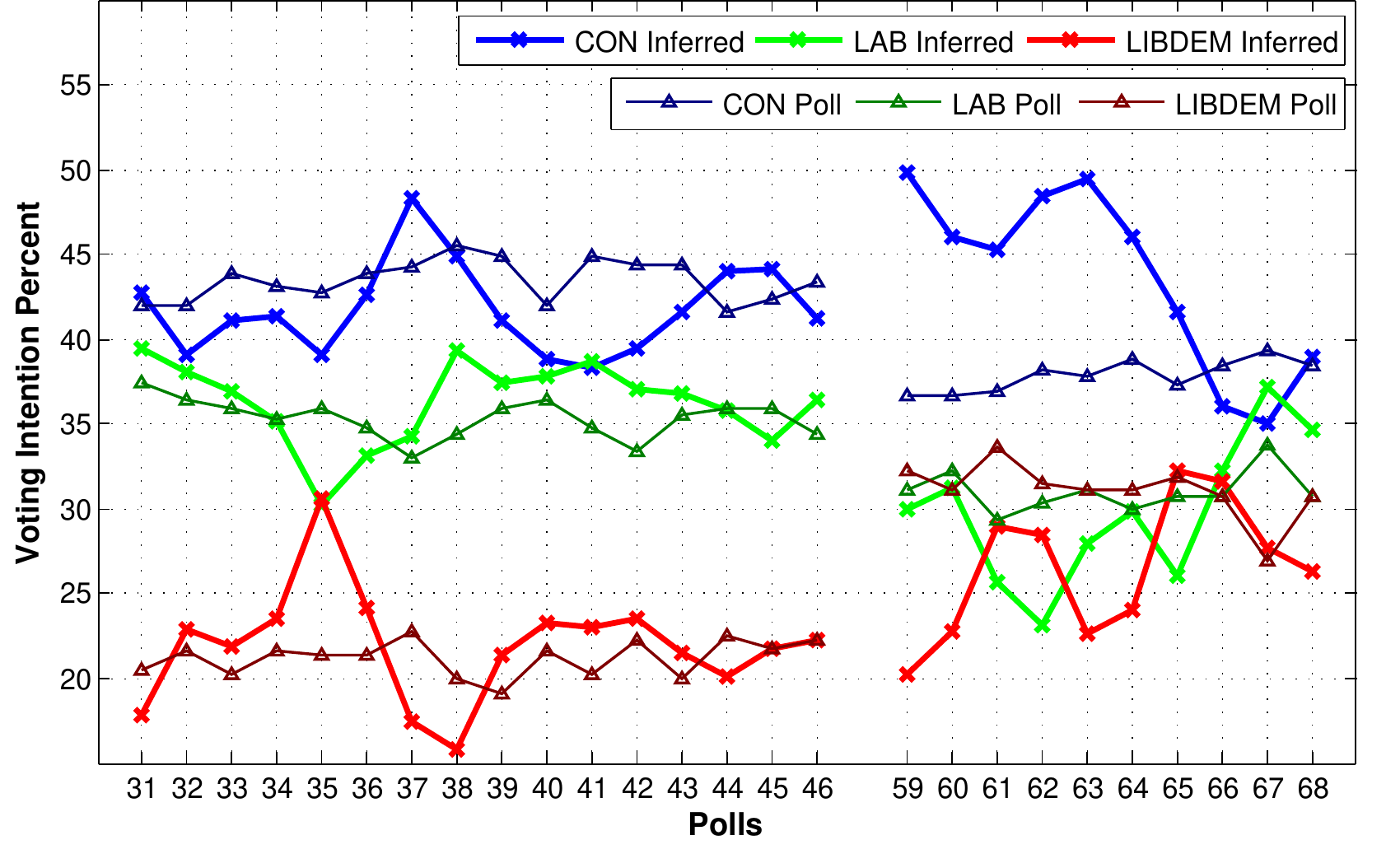}
\caption{SPOSW under thresholded DCS.}
\label{fig_vi_example_DCS_SPOSW}
\end{figure}

Inference results for MTS and DCS are presented in Tables \ref{table_vi_MTS} and \ref{table_vi_DCS} respectively. Again, it becomes apparent that combining SentiWordNet with POS tagging and extending tweets with WordNet's core senses (SPOSW) gives out the best inference performance as well as that DCS performs on average better than MTS. SnPOS and SPOS not only show a fairly bad performance in terms of MAE and MRE, but also do not deliver statistically significant results. On the contrary, SPOSW's inferences are shown to be statistical significant; its best performance is now achieved under DCS by performing thresholding reaching an MAE of $3.49 \pm 2.98$ and an MRE of $0.0962$ (see Figure \ref{fig_vi_example_DCS_SPOSW}).

\section{Related work and further discussion of the results}
\label{section_voting_intention_discussion}
The topic of voting intention or electoral result inference from the content of Twitter is quite new in the scientific literature. Tumasjan \etal have published a paper, which after providing proof that Twitter is a platform where political discussions are indeed conducted, proposes a method for predicting the result of the German elections in 2009 \cite{tumasjan2010predicting}. Their method uses the Linguistic Inquiry and Word Count (LIWC2007) for semantic analysis \cite{Pennebaker2007a}; this tool produces a set of 12 dimensions -- not just positive or negative sentiment -- and therefore Tumasjan \etal introduced an averaging technique to acquire 1-dimensional representations. Their method like ours uses sets of keywords (party and politician names) to select politically oriented tweets and then proposes a model for matching the Twitter traffic per party to the final result of the elections. Another paper \cite{OConnor2010} presents a different method for tracking voting intention polls based on the ratio of positive versus negative sentiment per tweet; their learning approach has some similarities to ours (it is based on linear regression), but only deals with bivariate problems, \ie polls with two outcomes. Nevertheless, a paper published after indicated that the methods in \cite{tumasjan2010predicting} and \cite{OConnor2010} were problem specific and not very generic, as their application was proven to not do better than chance in predicting the result of the US congressional elections in 2009 \cite{Gayo-Avello2011}. Another paper showed that the popular tool of Google Trends has a limited capacity in predicting election results \cite{Lui2010}.

Moreover, an interesting paper conducted a further analysis of the aforementioned methodologies and proposed a triplet of necessary standards that any theory aiming to provide a consistent prediction of elections using content from the Social Media should follow \cite{Metaxas2011a}. The authors recommend that firstly the prediction theory should be formed as a well defined algorithm, secondly the analysis should be aware of the different characteristics arising in the Social Web and thirdly that the method must include experimental justification on why it works.

Based on those propositions, we tried to formalise the steps in our methodology, we provided statistical significance figures for our results and finally by forming the three different schemes (SnPOS, SPOS and SPOSW) for extracting sentiment from tweets, we tried to encounter some of the special characteristics of Twitter's textual stream. Particularly in SPOSW, where we enriched tweets by adding synonymous words in order to enhance their semantic interpretation, we observed a significant improvement on inference performance.

Similarly to the aforementioned works, the first two sentiment extraction methods that we tried (SnPOS and SPOS) showed poor performance and were not statistically significant. On average, modelling sentiment with DCS performed better than MTS. This is quite interesting as -- by definition -- DCS compresses more information than MTS; one might argue that this compression results in a much clearer signal, \ie removes some noisy observations, especially when applied on extended tweets. The contribution of thresholding in our experiments is ambiguous; sometimes it improves the inference performance and in most occasions reduces the MRE. As we mentioned in the beginning, the optimal value as well as the contribution of thresholding could be investigated further.

We remind the reader that in this report the presented methods for modelling voting intention polls are preliminary. Several aspects are a matter of future research. Primarily, our methods should be applied to other data sets as well, to come up with experimental proof about their capability to generalise. Another important factor is the choice of keywords that are used to select tweets relevant to the task at hand. We argued that those keywords can be selected from topic-related repositories; still, the influence of each keyword should be modelled. Ideally, an automatic mechanism or algorithm should be compiled in order not only to select an optimal subset of keywords, but also to quantify the contribution of each term. Moreover, possible biases introduced by the use of different sentiment analysis tools should also be considered; tools that incorporate special emotional expressions used in instant messaging and Twitter such as `;-)' or `:D' might achieve a better performance.

A general conclusion that could be extracted or reconfirmed from the presented work is that Social Media do encapsulate content related to political opinion; extending tweets with synonymous terms probably assists in the amplification of this signal. However, it is very important to contemplate that the validity of voting intention polls is questionable. Are those polls a good representation of what is actually happening? Usually, polls from different agencies tend to have significant differences in their percentages and quite often those differences exceed, for example, the lowest total MAE derived in our experimental process (3.49\%). Therefore, the formation of a consistently good ground truth is another issue that approaches like this one must resolve.

\bigskip
\bigskip

\section*{Acknowledgements}
{\small Vasileios Lampos is partially supported by EPSRC (DTA/SB1826). Lampos would like to thank Nick Fyson for his significant help in shaping the lists with the most influential politicians (or keywords) per political party. He is also grateful to Prof. Nello Cristianini for providing some useful feedback on this preliminary work.}

\bigskip
\bigskip

% Bibliography
\bibliographystyle{acmlarge}
\bibliography{library}

\bigskip
\bigskip

\appendix

\section{Appendix: Search Terms}
\label{Ap:voting_intention}
Terms used to select tweets related to the major political parties in the UK. Character `\_' denotes that an empty space and `\#' denotes a Twitter topic. Tables \ref{table_conservatives}, \ref{table_labour} and \ref{table_libdem} hold the terms for the Conservative (52 terms), Labour (57 terms) and Liberal Democrat party (62 terms) respectively.

\bigskip

\begin{table}[h!]
\renewcommand{\arraystretch}{1.2}
\centering
\tbl{Terms for the Conservative Party (\textbf{CON}) -- 52 terms}{
\(\begin{tabular}{cccc}
\hline
\#conservative       & CONSERVATIVE          & Eric Pickles      & Margaret Thatcher\\
\#Tories             & Conservative          & Francis Maude     & Michael Gove\\
\#tories             & Conservative Party    & George Osborne    & Oliver Letwin\\
\#Tory               & Crispin Blunt         & Gerald Howarth    & Owen Paterson\\
\#tory               & David Cameron         & Grant Shapps      & Patrick McLoughlin\\
\_TORIES             & David Davis           & Hugo Swire        & Philip Hammond\\
\_tories             & David Mundell         & Iain Smith        & President Cameron\\
Alan Duncan          & David Willetts        & Jeremy Hunt       & Theresa May\\
Andrew Lansley       & debate Cameron        & John Redwood      & Tories\\
Andrew Mitchell      & Dominic Grieve        & Ken Clarke        & vote Cameron\\
Boris Johnson        & Duncan Smith          & Lady Warsi        & vote conservative\\
Caroline Spelman     & Edward Leigh          & Leader Cameron    & vote Tory\\
Cheryl Gillan        & Edward Vaizey         & Liam Fox          & William Hague\\
\hline
\end{tabular}\)
}
\label{table_conservatives}
\end{table}

\bigskip

\begin{table}[h!]
\renewcommand{\arraystretch}{1.2}
\centering
\tbl{Terms for the Labour Party (\textbf{LAB}) -- 57 terms}{
\(\begin{tabular}{cccc}
\hline
\#labour            & debate Brown          & LABOUR                & Rosie Winterton\\
\_Ed \_Balls        & Diane Abbott          & Labour                & Sadiq Khan\\
Alan Johnson        & Douglas Alexander     & Labour Party          & Shaun Woodward\\
Alastair Campbell   & Ed Miliband           & Leader Brown          & Stella Creasy\\
Andy Burnham        & Gordon Brown          & Liam Byrne            & Tessa Jowell\\
Angela Eagle        & Government Brown      & Luciana Berger        & Tom Harris\\
Ann McKechin        & Harriet Harman        & Maria Eagle           & Tom Watson\\
Ben Bradshaw        & Hilary Benn           & Mary Creagh           & Tony Blair\\
Caroline Flint      & Ivan Lewis            & Meg Hillier           & Tony Lloyd\\
Charles Clarke      & Jim Murphy            & MILIBAND              & vote Brown\\
Chris Bryant        & John Denham           & Miliband              & vote labour\\
Chuka Umunna        & John Healey           & Peter Hain            & Yvette Cooper\\
David Blunkett      & John Prescott         & Peter Mandelson       & \\
David Lammy         & Jon Trickett          & President Brown       & \\
David Miliband      & Kerry McCarthy        & Prime Minister Brown  & \\
\hline
\end{tabular}\)
}
\label{table_labour}
\end{table}

\bigskip

\begin{table}[ht!]
\renewcommand{\arraystretch}{1.2}
\centering
\tbl{Terms for the Liberal Democrat Party (\textbf{LIB}) -- 62 terms}{
\(\begin{tabular}{cccc}
\hline
\#Clegg                  & Fiona Hall       & LIBERAL               & Norman Backer\\
\#clegg                  & Greg Mulholland  & Liberal               & Norman Lamb\\
\#liberal                & Jenny Willott    & Liberal Democrat      & Paul Burstow\\
\_Ed Davey               & Jeremy Browne    & Lord McNally          & President Clegg\\
\_Jo Swinson             & John Hemming     & Lord Shutt            & Sarah Teather\\
Alistair Carmichael     & John Pugh         & Lorely Burt           & Simon Hughes\\
Andrew Stunell          & John Thurso       & Lynne Featherstone    & Steve Webb\\
Charles Kennedy         & Julian Huppert    & MP Clegg              & Tessa Munt\\
Chris Huhne             & Leader Clegg      & Mark Hunter           & Tim Farron\\
Dan Rogerson            & Lembit Opik       & Martin Horwood        & Tom McNally\\
Danny Alexander         & Lib\_ \_Dem       & Michael Crockart      & Vince Cable\\
David Heath             & LIBDEM            & Mike Crockart         & Vincent Cable\\
David Shutt             & LibDem            & Mike Hancock          & vote Clegg\\
debate Clegg            & Libdem            & Nicholas Clegg        & vote liberal\\
Duncan Hames            & libDem            & Nick Clegg            &\\
Edward Davey            & libdem            & Nick Harvey           &\\
\hline
\end{tabular}\)
}
\label{table_libdem}
\end{table}

\end{document}